\documentclass[aps,prapplied,reprint,superscriptaddress]{revtex4-2}
\usepackage{graphicx}
\usepackage{dcolumn}
\usepackage{bm}
\usepackage[utf8]{inputenc}
\usepackage[T1]{fontenc}
\usepackage{booktabs, array, mathptmx, float, tabularx, booktabs, lipsum, multirow, amstext}
\usepackage{amsmath}
\usepackage{mathptmx}
\usepackage{mathrsfs}
\DeclareMathAlphabet{\mathcal}{OMS}{cmsy}{m}{n}
\usepackage{siunitx, xcolor}
\usepackage[version=4]{mhchem}

\begin{document}

\title{Robust high-frequency laser phase noise suppression by adaptive Pound-Drever-Hall Feedforward}

\author{Yu-Xin Chao}
\thanks{These authors contributed equally to this work.}
\affiliation{State Key Laboratory of Low Dimensional Quantum Physics, Department of Physics, Tsinghua University, Beijing 100084, China.}

\author{Zhen-Xing Hua}
\thanks{These authors contributed equally to this work.}
\affiliation{State Key Laboratory of Low Dimensional Quantum Physics, Department of Physics, Tsinghua University, Beijing 100084, China.}

\author{Xin-Hui Liang}
\affiliation{State Key Laboratory of Low Dimensional Quantum Physics, Department of Physics, Tsinghua University, Beijing 100084, China.}

\author{Zong-Pei Yue}
\affiliation{State Key Laboratory of Low Dimensional Quantum Physics, Department of Physics, Tsinghua University, Beijing 100084, China.}

\author{Chen Jia}
\affiliation{State Key Laboratory of Low Dimensional Quantum Physics, Department of Physics, Tsinghua University, Beijing 100084, China.}

\author{Li You}
\affiliation{State Key Laboratory of Low Dimensional Quantum Physics, Department of Physics, Tsinghua University, Beijing 100084, China.}
\affiliation{Frontier Science Center for Quantum Information, Beijing, China.}
\affiliation{Hefei National Laboratory, Hefei, Anhui 230088, China.}

\author{Meng Khoon Tey}
\email{mengkhoon\_tey@mail.tsinghua.edu.cn}
\affiliation{State Key Laboratory of Low Dimensional Quantum Physics, Department of Physics, Tsinghua University, Beijing 100084, China.}
\affiliation{Frontier Science Center for Quantum Information, Beijing, China.}
\affiliation{Hefei National Laboratory, Hefei, Anhui 230088, China.}
\date{\today}

\begin{abstract}
Suppressing high-frequency laser phase noise, particularly at frequencies near and beyond typical feedback bandwidths of a few MHz, is a critical yet challenging task in many advanced  applications. Feedforward-based methods generally outperform feedback in high-frequency range, but their performances are more susceptible to perturbations. In this work, we focus on the Pound-Drever-Hall (PDH)-feedforward method we demonstrated recently [Yu-Xin Chao et al., Optica 11(7), 945-950 (2024)] and analyze the factors that affect its long-term stability. By constructing a simple circuit allowing for adaptive control of the feedforward gain in response to power fluctuations of cavity transmission, we demonstrate a robust $\geq 40$~dB suppression of laser phase noise around 2~MHz and a noise suppression bandwidth up to 50~MHz. In comparison, when using normal PDH feedback, robust noise suppression of over 40 dB can only occur for frequencies below tens of kHz in most setups. Our findings may pave the way for general usage of PDH feedforward and allow for simple construction of low-noise lasers for precise quantum controls and precision metrology.
\end{abstract}

\maketitle

\section{Introduction}

Frequency-stabilized low-noise lasers are the heart to many state-of-the-art technologies, including gravitational wave sensing~\cite{2015_LIGO_AdvancedLIGO}, optical clocks~\cite{2015_RMP_OpticalAtomicClock,2021_Katori_portableClock,2019_YeJun_SrClock,2019_Leibrandt_AlClock,2021_LinYiGe_SrClock,2022_ChangHong_OpticalLatticeClock,2024_Pan_SrOpticalLatticeClock}, ultra-low noise photonic microwave generation~\cite{2011Fortier_Optical_microwave,2017XieMicrowave,2020Quinlan_Science_Microwave,2020Kippenberg_photonicMicrowave}, quantum computation with atomic qubits~\cite{2018_Browaeys_Coherence,2019_Levine_MultiqubitGate,2020_ManuelEndres_AKEA,2019_APR_Ion,2021_RMP_Ion,2022_Zhan_Rydberg_gate,2022_Shao_PhaseGate,2023_Lukin_ParallelGate,2023_Sun_OffResonant,2024_Bernien_DualSpecies} and coherent synthesis of ultracold molecules~\cite{2008Ni_molecules,2014_Hanns_RbCs,2014_Cornish_RbCs,2016_WangDaJun_NaRb,2020_Silke_NaK,2021_LuoXinYu_NaK,2022Bo_molecules}. Behind all these lasers lies a frequency-discriminating and locking method called the Pound-Drever-Hall (PDH) technique developed about 40 years ago. This method compares the frequency of a laser with a Fabry-P$\rm{\acute{e}}$rot cavity's resonance, and converts their difference into an electrical signal for high-speed feedback~\cite{1983_PDH_original,PDH_Black}. Combined with ultra-stable cavities, PDH feedback has become a standard practice for building narrow-linewidth lasers.

However, as the name implies, any feedback mechanism inherently involves a response delay between the corrective action and the occurrence of an error. This delay sets an upper limit on the feedback bandwidth, $f_b$, beyond which the noise is not only unmitigated but can even be exacerbated by the feedback. As a rule of thumb, noise at a Fourier frequency $f_n$ can only be suppressed by a factor of $\left| 1+G_{\rm fb}(f_n)\right| ^2 \sim (f_b/f_n)^2$ (in terms of noise power) with feedback, where $G_{\rm fb}(f_n)$ represents the closed-loop voltage gain~\cite{2003_Fox_StabilizingDiodeLaser}. In most PDH setups, $f_b$ is limited to a few MHz~\cite{1984_Hall_DelayLineEOM,2018_Endo_ResidualPhaseNoiseEOM,2022_Palmer_IntracavityEOM,2021_Polzik_HighFrequency}. Consequently, substantial noise suppression exceeding 40~dB is achievable only for frequencies below tens of kHz, while noise in the MHz range is not suppressed but rather amplified.

In comparison, feedforward-based methods are more effective at addressing high-frequency laser phase noise~\cite{2009_Hossein_MZI,2012_Hossein_MZI,2016_Tetsuya_TrackingInterferometer,2016_Barry_FrequencyComb,2017_Lintz_Note,2015Scharnhorst_Feedforward,2016Xu_FeedforwardwithComb,2019_HeZuyuan_AOM,2022_Demarco_CavityRef,2024_Chao_PDHFeedForward}. In essence, these methods employ an optical fiber to delay the light so as to wait for the correction signal, ensuring that phase errors and their corrective signals arrive at the phase-correcting actuator simultaneously. As a result, a well-implemented feedforward system serves as a real-time remedy for any phase errors, thus offering a significantly higher bandwidth compared to feedback methods.

\begin{figure*}[!ht]
	\centering
	\includegraphics[width=1.9\columnwidth]{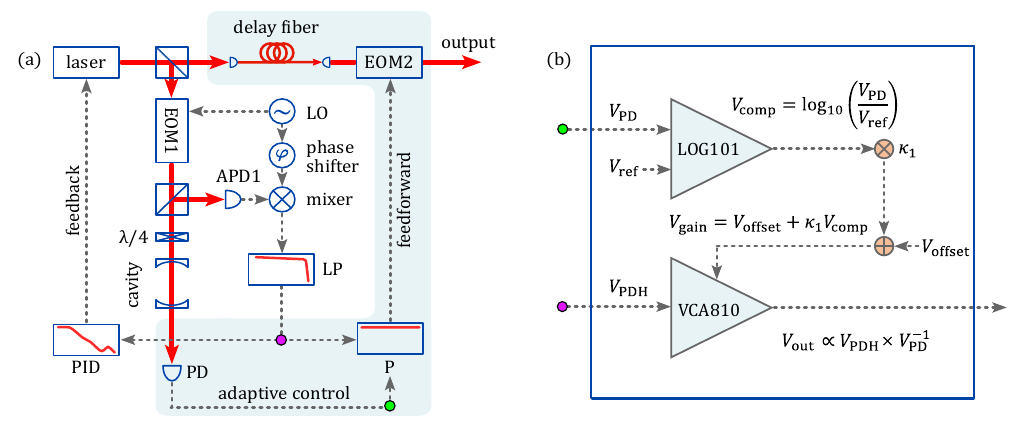}
	\caption{
		Working principles of PDH feedforward with adaptive gain control. (a) A typical PDH feedback setup (white background) for stabilizing laser's frequency is used in conjunction with a feedforward add-on (blue shaded area) for suppressing high-frequency phase noise. When the laser is feedback-locked to a cavity resonance, PDH feedforward is accomplished simply by applying the residual PDH error signal to an electro-optic modulator (EOM2) after a delay fiber with properly compensated length. The transmission power through the cavity is monitored by a photodiode (PD) and is used to control the gain of the feedforward signal through the loop filter P. LO, local oscillator. LP, low-pass filter. (b) The design principle of the adaptive loop filter P which outputs a feedforward signal proportional to $V_{\rm PDH}(t)/V_{\rm PD}(t)$, see text.
	}
	\label{Principle}
\end{figure*}

Previously, several techniques have been employed to extract high-frequency laser phase noise for feedforward applications. They include delayed self-homodyne measurement with an unbalanced Mach-Zehnder interferometer (MZI)~\cite{2009_Hossein_MZI,2012_Hossein_MZI,2016_Tetsuya_TrackingInterferometer,2016_Barry_FrequencyComb,2017_Lintz_Note,2017_HuangWei_CoherentEnvelope,2019_ZhanMingSheng_EnvelopRatio,2021_Polzik_HighFrequency} and heterodyne detection by beating a light field and its transmission through a filtering cavity~\cite{2022_Demarco_CavityRef}. In a recent development, we have demonstrated that the classic PDH technique can be extended to implement feedforward effectively~\cite{2024_Chao_PDHFeedForward}. In practice, this is achieved by sending the residual PDH error signal (of a laser locked to a cavity) to a phase-correcting electro-optic modulator (EOM) placed behind a delay fiber (Fig.~\ref{Principle}(a)). We showed that this method enables a noise suppression exceeding 37~dB at 2~MHz, outperforming usual PDH feedback by four orders of magnitude, and extends the noise-suppression bandwidth to tens of MHz~\cite{2024_Chao_PDHFeedForward}. This method has recently been utilized to significantly enhance the STIRAP transfer efficiency of ultracold RbCs molecules~\cite{2024_Cornish_STIRAP}.

However, like all feedforward-based methods, the performance of PDH feedforward is vulnerable to environmental perturbations. This instability arises because feedforward is inherently an open-loop process, offering only one-off compensation in contrast to feedback. As a rule of thumb, the noise suppression power-ratio of feedforward is given by $ \left| 1-G_{\rm ff}(f_n) \right| ^{-2}$, where $G_{\rm ff}(f_n)$ represents the overall voltage gain of the feedforward loop. To reach a 40~dB phase noise suppression, the deviation of $G_{\rm ff}(f_n)$ from unity must be less than $1\%$. If $G_{\rm ff}(f_n)$ drifts by $3\%$, the suppression degrades to 30~dB; a $10\%$ offset reduces noise suppression to 20~dB. These vulnerabilities make feedforward methods less prevalent than feedback methods. For PDH feedforward, $G_{\rm ff}(f_n)$ can be affected by several factors, including light power fluctuations, mechanical drift that deteriorates the coupling efficiency of light into the cavity, and temperature variations affecting the response of electronic circuitry, etc.

In this work, we demonstrate a simple method which significantly enhances the robustness and performance of the PDH feedforward technique. More explicitly, we constructed an electronic circuit which dynamically compensates for the feedforward gain in response to variations in the optical power transmitted through the cavity. Implemented on a standard laser setup, this design demonstrates a long-term phase-noise suppression capability exceeding 40~dB at a noise frequency of 2~MHz and extends the phase noise suppression bandwidth to 50~MHz.

This article is organized as follows. Section~\ref{sec:PDH_feedforward_and_vulnerabiilits} explains the working principles of PDH feedforward and then analyzes its vulnerabilities by examining the PDH error signal. Section~\ref{sec:design_principle} describes the primary concept of our adaptive-gain solution. Section~\ref{sec:detection_and_optimization} discusses the methods used to probe and optimize the performance of this feedforward scheme. The performance of the scheme and its reliability are discussed in Sec.~\ref{sec:performance_and_reliability}. In Sec.~\ref{sec:advantages}, we compare the benefits of this adaptive-gain design with the power stabilization technique and discuss potential extensions of this adaptive method. The conclusion is presented at the end.

\section{Overview of the technique}

\subsection{PDH feedforward and its vulnerabilities}\label{sec:PDH_feedforward_and_vulnerabiilits}

The schematic diagram of Fig.~\ref{Principle}(a) demonstrates how to apply both feedback and feedforward to suppress laser phase noise using the PDH technique. The primary part of the construction is the familiar PDH feedback (white background in Fig.~\ref{Principle}(a)). Implementing PDH feedforward requires only a few supplemental components: a delay fiber, an EOM for correcting phase error, and a loop filter with frequency-independent proportional (P) gain (as depicted within the blue shaded region in Fig.~\ref{Principle}(a)). Notably, the photodetector (PD) used for detecting the transmission power through the cavity is absent in the original PDH feedforward scheme~\cite{2024_Chao_PDHFeedForward}.

{Contents of the PDH error signal} --- When a laser is feedback-locked to a cavity, its electric field can be expressed by $E_0 \exp \left\{i\left[ \omega_c t + \phi(t)\right]\right\}$. Here, $E_0$ is the field amplitude, $\omega_c$ denotes the resonant frequency of the cavity, and $\phi(t)$ represents the instantaneous residual phase error at time $t$. In the PDH setup depicted in Fig.~\ref{Principle}(a), EOM1 imprints a sinusoidal phase modulation of $\beta \sin(\Omega t)$ to the light incident to the cavity, resulting in an incident field consisting of many frequency components:
\begin{equation} \label{eq_incident_bessel}
	E_{\rm in}(t)= E_0 \sum_k J_k(\beta) \exp \left\{i\left[(\omega_c + k\Omega) t+ \phi (t)\right] \right\}.
\end{equation}
Here $J_k(x)$ represents the $k^{\rm th}$ order Bessel function of the first kind ($k=0, \pm 1, \pm 2, ...$).
Generally, when a laser is properly feedback-locked to a cavity, its typical linewidth should be well below $\sim0.01\gamma_c$ (where $\gamma_c$ represents the cavity's FWHM linewidth, usually around 10~kHz for high-finesse cavities). In most PDH applications, the characteristic frequency of unmitigated phase noise is around the typical feedback bandwidth of a few hundred kHz to a few MHz. This frequency range is much higher than the cavity linewidth $\gamma_c$ but much lower than the typical PDH modulation frequency $\Omega/2\pi$, which is on the order of tens of MHz. Given these well-separated frequency scales, it can be assumed that all of the field components in Eq.~(\ref{eq_incident_bessel}), except for the carrier at $\omega_c$, are completely reflected by the cavity. Consequently, the field reflected from the cavity can be approximated by
\begin{equation} \label{eq_reflected}
	E_{\rm ref}(t)= -E_{\rm in}(t)  + E_0^\prime J_0(\beta) \exp(i\omega_c t),
\end{equation}
where the negative sign indicates a phase shift of $\pi$ acquired by the reflected fields and $E_0^\prime$ represents the field amplitude coupled into the cavity. Under this simplified model, the PDH error signal is given by~\cite{2024_Chao_PDHFeedForward}
\begin{equation} \label{eq_PDH_error}
	V_{\rm PDH}(t) \propto (E_0^{\prime})^2 J_0(\beta) J_1(\beta) \sin\left[\phi(t)\right].
\end{equation}
Equation~(\ref{eq_PDH_error}) signifies that (i) the PDH signal is directly proportional to the instantaneous phase noise $\phi(t)$ when $|\phi(t)|\ll 1$; (ii) its amplitude is proportional to the power of the field which is coupled into the cavity.

When performing feedforward noise cancellation, we simply imprint a phase shift proportional to $V_\mathrm{PDH}(t)$ to the laser's light field using EOM2. This results in an output field of $E_{\rm out} = E_1 \exp\left\{i \left[ \omega_c t + \phi(t)- G_{\rm ff} \sin\phi(t^{\prime})\right]\right\}$. The notation $t^\prime$ is used to emphasize the potential temporal mismatch between the phase error and the corrective signal. When $|\phi(t)|\ll 1$, the phase error $\phi(t)$ can be effectively compensated by setting a constant $G_{\rm ff}=1$ for all frequencies, and matching $t$ and $t^\prime$.

{Vulnerabilities --- According to Eq.~(\ref{eq_PDH_error}) however, $V_{\rm PDH}(t)$, and thereby $G_{\rm ff}$, is directly proportional to light power coupled into the cavity. Therefore, fluctuations in the incident light power or drifts in the light-cavity coupling efficiency can lead to deviation of $G_{\rm ff}$ from the ideal value of $1$. This issue is the primary source of instability in PDH feedforward. Other secondary effects are mostly due to temperature drift, which can change the sensitivities of photo-detectors, the modulation effect of EOMs ($V_{\rm PDH}$ is affected by the phase-modulation amplitude $\beta$), and the gain of electronic circuits, etc.

\subsection{PDH feedforward with adaptive gain control}\label{sec:design_principle}

This work aims to improve the robustness of the PDH feedforward technique by tackling its primary instability caused by variations in the optical power coupled into the cavity. Instead of stabilizing the power transmitted through the cavity, we introduce a more compact and reliable solution of redesigning the loop filter P in Fig.~\ref{Principle}(a) so that it can adapt to the changes in the cavity transmission power, thus maintaining a constant $G_{\rm ff}$. This approach offers several additional advantages over the power stabilization method, which we will discuss in depth later.

The design principle of loop filter P is straightforward. Since the PDH error signal $V_{\rm PDH}(t)$ after the demodulation mixer is proportional to the instantaneous cavity transmission power $P(t)$ (or equivalently the output voltage of the PD after the cavity $V_{\rm PD}(t)$), achieving an overall feedforward gain $G_{\rm ff}$ independent of fluctuations in $P(t)$ requires the output of loop filter P to be of the form:
\begin{equation}\label{eq_main_idea}
	V_{\rm out}(t) = G \frac{P_0}{P(t)}V_{\rm PDH}(t) = G \frac{ V_{\rm ref}}{V_{\rm PD}(t)} V_{\rm PDH}(t).
\end{equation}
Here, $P_0$ ($V_{\rm ref}$) represents the reference transmission power (PD voltage) at which $G$ is adjusted to give $G_{\rm ff}=1$. We realize Eq.~(\ref{eq_main_idea}) using the circuit illustrated in Fig.~\ref{Principle}(b) which operates as follows. First, $V_{\rm PD}(t)$ is compared with a reference voltage using a logarithmic amplifier (LOG101, Texas Instruments), producing an output signal
\begin{eqnarray} \label{eq_LOG101_output}
	V_{\rm comp}(t) &=& V_0 \log_{10}\left[\frac{V_{\rm PD}(t)}{V_{\rm ref}}\right],
\end{eqnarray}
where $V_0=1\rm{V}$. $V_{\rm comp}$ is subsequently multiplied by a factor $\kappa_1$ and offset by $V_{\rm offset}$ to give $V_{\rm gain} = V_{\rm offset} + \kappa_1 V_{\rm comp}$. This $V_{\rm gain}$ is finally used to control the amplification factor of $V_{\rm PDH}(t)$ via a voltage-controlled amplifier (VCA810, Texas Instruments) whose gain increases exponentially with respect to $V_{\rm gain}$. Overall, these procedures give an output signal of
\begin{eqnarray} \label{eq_VCA810_output}
	V_{\rm out}(t) &=& 10^{(\kappa_2 V_{\rm gain} + \kappa_3)}\times V_{\rm PDH}(t) \nonumber\\
	&=& 10^{(\kappa_2 V_{\rm offset} + \kappa_3)} \left[\frac{V_{\rm PD}(t)}{V_{\rm ref}}\right]^{\kappa_1 \kappa_2 V_0}V_{\rm PDH}(t),
\end{eqnarray}
where $\kappa_2$ and $\kappa_3$ are intrinsic parameters of VCA810. To conform with Eq.~(\ref{eq_main_idea}), $\kappa_1$ is set to $-1/\kappa_2 V_0$ using the actual circuitry depicted in Appendix~\ref{appendix_circuit_details}. During optimization, $V_{\rm offset}$ is used to adjust $G = 10^{\kappa_2 V_{\rm offset} + \kappa_3}$ to give $G_{\rm ff}=1$.

\subsection{Phase noise detection and optimizing procedures of feedforward} \label{sec:detection_and_optimization}
We probe the laser phase noise using heterodyne spectrum obtained by beating the frequency-shifted output light after EOM2 and the cavity-transmitted light on an APD (see Appendix~\ref{appendix_setup_details}). The heterodyne spectrum obtained this way reflects the true noise spectrum in the output light when the noise frequencies are much higher than the cavity linewidth $\gamma_c$ ($\approx 14.5$~kHz for our setup)~\cite{2024_Chao_PDHFeedForward,2019_Udem_SimpleMeasurement}. To measure the noise suppression capability of feedforward at a particular frequency $f_{\rm n}$, we first inject a weak sinusoidal phase modulation into the light field (see Appendix~\ref{appendix_setup_details}). The weak injection does not produce discernible changes in the noise spectrum of the feedback-locked laser, except for the appearance of a pair of injected sidebands, as indicated by the black circles in Figs.~\ref{Bandwidth}(a) and (b). We then compute an attenuation parameter $A_{\rm heterodyne}(f_{\rm n}) \equiv P_{\rm w}(f_{\rm n})/P_{\rm w/o}(f_{\rm n})$, which is the ratio of the beat note powers $P_{\rm w(w/o)}$ of the injected sidebands with (without) feedforward.

Once the laser is feedback-locked to the cavity, we proceed to  optimize the feedforward using the following procedures. First, we set the cavity transmission power to $P_0$ (which gives a $V_{\rm PD}$ about half of the PD's saturation level), and then adjust $V_{\rm ref}$ to match the reference $V_{\rm PD}$ such that $V_{\rm comp}=0$. Next, we inject a weak sinusoidal phase modulation at 2~MHz into the light field. We then fine-tune $V_{\rm offset}$ and simultaneously adjust the length of the coaxial cable connecting loop filter P and EOM2~\cite{HowToMatchDelay} to obtain a minimal $A_{\rm heterodyne}$. Finally, we switch on the adaptive-gain control and then adjust $\kappa_1$ to minimize $A_{\rm heterodyne}$ over a wide range of cavity-transmission powers. Notably, although $\kappa_1$ should ideally be set to $-1 / \rm{\kappa_2 V_0}$ as discussed in Sec.~\ref{sec:design_principle}, we realized that the most efficient approach is to optimize $\kappa_1$ according to the observed performance. This is to account for the inevitable nonlinear responses in the system, primarily arising from operating electronic devices, such as photo-detectors, mixers, and amplifiers, near their output limits. 

\section{Results and discussions}

\begin{figure}[!ht]
	\centering
	\includegraphics[width=0.99\columnwidth]{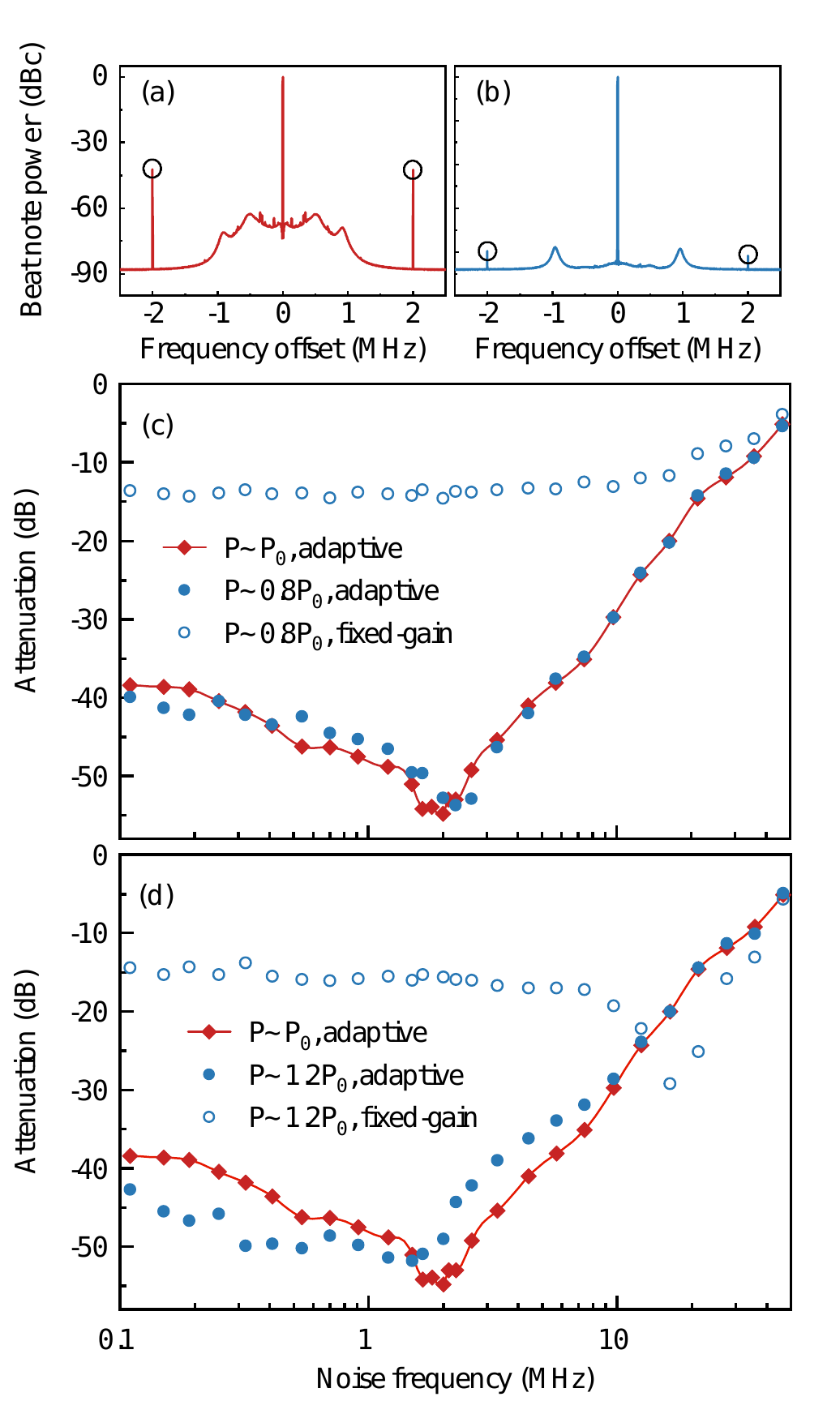}
	\caption{
		Phase noise suppression capability of PDH feedforward with and without adaptive-gain control. (a) and (b), heterodyne spectra obtained by beating the frequency-shifted light after EOM2 with the cavity-filtered light on APD2 (see Fig.~\ref{fig_DetailedSetup} for illustration), before (a) and after (b) applying PDH feedforward. The data are recorded with a resolution bandwidth of 50~Hz. The black circles highlight the artificially injected sinusoidal phase noise. The remaining noise bumps around 1~MHz in (b) are due to laser intensity noise which cannot be suppressed by PDH feedforward. (c) and (d), measured noise attenuation $A_{\rm heterodyne}$ as a function of injected noise frequency $f_n$. The feedforward performance is optimized at $f_n=2$~MHz and at the cavity transmission power $P_0$. At the optimized conditions, the maximum attenuation reaches $-54$~dB with adaptive-gain control (red diamonds) activated. The transmission power is set to 80\% of $P_0$ in (c) and 120\% of $P_0$ in (d). For all measurements, the beat note powers of the injected sidebands are set to around -40~dBc before applying feedforward.
	}
	\label{Bandwidth}
\end{figure}

\subsection{Performance and reliability}\label{sec:performance_and_reliability}

Figures~\ref{Bandwidth}(c) and \ref{Bandwidth}(d) exhibit the feedforward performance ($A_{\rm heterodyne}$) of our setup, measured at various noise frequencies and cavity transmission powers, with and without adaptive-gain control. The red diamonds represent the results obtained with the cavity transmission power set at $P_0$ (at which the system is optimized) and with the adaptive gain control activated. The data show a remarkable attenuation of over $-40$~dB from 250~kHz to 4.4~MHz, and a maximum attenuation of over -50~dB near the optimized frequency of 2~MHz. The decline around 2~MHz is caused by variations in the gain and the group delay of the electrical components with frequency~\cite{cable_length_comment}, while the feedforward bandwidth of about 50~MHz is primarily limited by the low-pass filter (LP) we used. Figures~\ref{Bandwidth}(c) and \ref{Bandwidth}(d) illustrate the scenarios where the cavity transmission power is set to 80~\% and 120~\% of $P_0$, respectively. In both cases, feedforward with fixed gain (empty blue circles) performs poorly, while the attenuation of adaptive feedforward (filled blue circles) closely aligns with the optimal results measured at $P_0$. 

To demonstrate the robustness of our design, we fix the noise frequency at 2~MHz and measure the attenuation parameter $A_{\rm heterodyne}$ as a function of the cavity transmission power. The results shown in Fig.~\ref{Stability} reveal that the attenuation of the adaptive feedforward remains over -40~dB when the cavity transmission varies from $0.5P_0$ to $1.5P_0$. In contrast, the attenuation of the fixed-gain feedforward decreases rapidly as the transmission power deviates from $P_0$. Notably, we find that adaptive-gain control helps improve the noise-suppression performance of PDH feedforward even at the optimized power level $P_0$, presumably because it also effectively corrects for fast light power fluctuations above kHz. 

\begin{figure}[htbp]
	\centering
	\includegraphics[width=0.98\columnwidth]{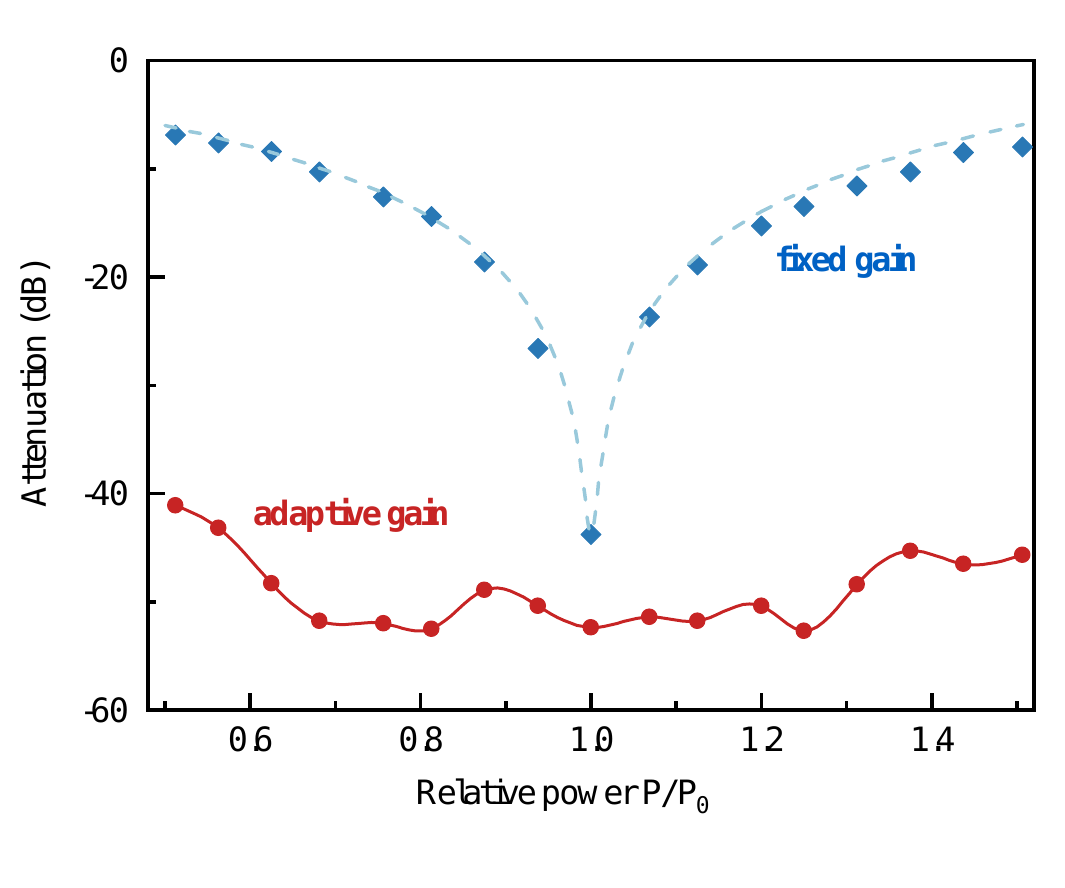}
	\caption{
		Phase noise suppression performance of PDH feedforward using fixed-gain and adaptive-gain as a function of cavity transmission power relative to $P_0$. The data are measured at noise frequency of 2~MHz. The blue dashed line represents theoretical predictions obtained by assuming the $G_{\rm ff}=1$ at $P=P_0$, and the non-zero value of $A_{\rm heterodyne,~fixed \mbox{-} gain}$ at $P=P_0$ is modeled using a stochastic white noise in the PDH signal.
	}
	\label{Stability}
\end{figure}

\begin{figure}[!ht]
	\centering
	\includegraphics[width=0.98\columnwidth]{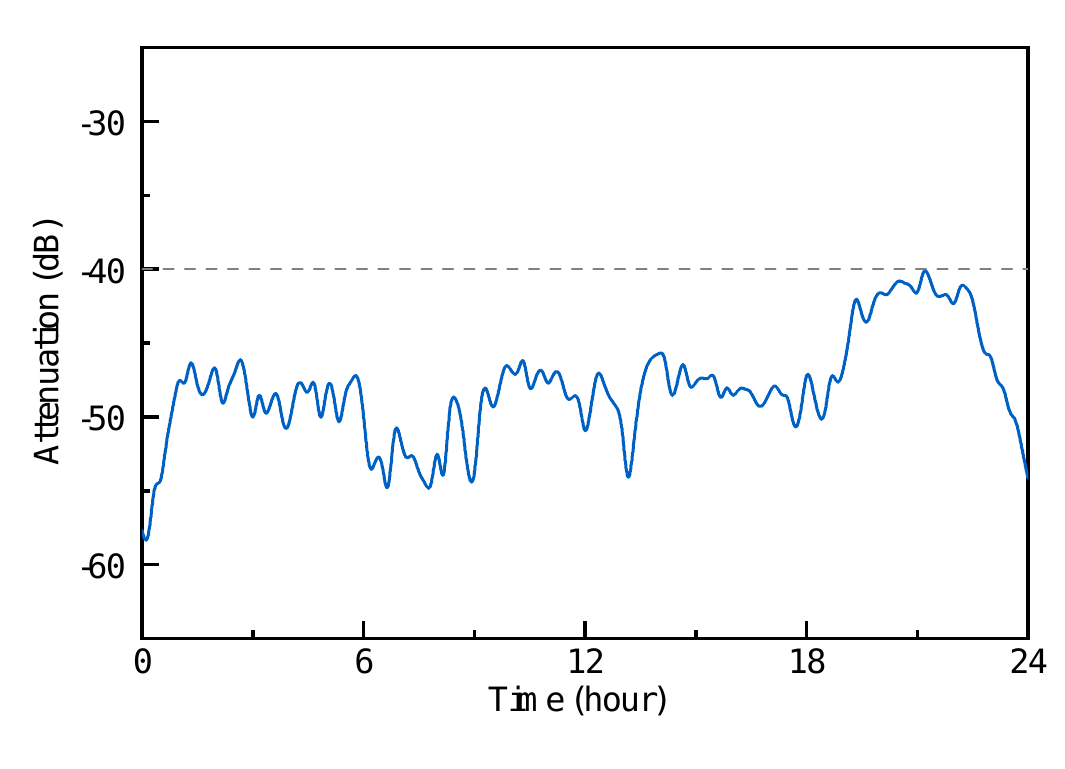}
	\caption{
		Noise suppression performance of the adaptive PDH feedforward system at 2~MHz, measured over a period of 24 hours (see text for conditions).
	}
	\label{LongTimeData}
\end{figure}

Beyond variations in cavity transmission power, other factors such as fluctuations in laboratory temperature and humidity can also degrade the stability of PDH feedforward. To verify the robustness of our design against all the environmental perturbations, we record the attenuation parameter at 2~MHz over a 24-hour period without making any adjustments to the system. Our setup exhibits a robust attenuation of over $-40$~dB, despite the fact that the laboratory temperature varies by 1.5 degrees Celsius and the cavity transmission fluctuates by 10~\% peak-to-peak around $P_0$ during this period. This result thereby confirms our claim that power fluctuation of the cavity transmission is the primary source of instability in PDH feedforward.

\subsection{Advantages of adaptive-gain control over power stabilization technique and future outlook}\label{sec:advantages}

The adaptive-gain control method demonstrated in this work offers several distinct advantages over the traditional method of stabilizing the light power transmitted through the cavity:
\begin{enumerate}
\item The adaptive-gain control can be fully integrated into the loop filter P using only a few chips. This makes it far more compact and reliable compared to the power stabilization method, which requires additional feedback circuits, a power actuator such as an acousto-optic modulator (AOM), and the actuator's driver.
\item The feedback loop for stabilizing cavity transmission can conflict with the feedback loop for stabilizing laser frequency, leading to degraded performance for both systems. In contrast, adaptive-gain control does not interfere with the PDH feedback loop, ensuring its proper operation in the presence of intensity noise.

\item As demonstrated in Fig.~\ref{Stability}, the adaptive scheme can easily handle a dynamic range of cavity transmission power much larger than what is typically achievable with the power stabilization scheme, unless a large power reserve is set aside.
\end{enumerate}

Outlook --- We observe that many factors can result in non-negligible deviation of $V_{\rm PDH}(t)$ from the ideal relation with $P(t)$ given by Eq.~(\ref{eq_PDH_error}). Such nonlinearities are not only due to operating an electronic device close to its saturation, but can also be caused by, for instance, changing the frequency offset between the laser and the cavity's resonance when using offset-locking~\cite{offsetlocking}. To handle variations of $V_{\rm PDH}(t)$ caused by multiple varying factors, one could replace the circuit section which generates $V_{\rm gain}$ (see Fig.~\ref{Principle}(b) and Fig.~\ref{fig_Circuitry}(a)) by a digital micro-controller. By pre-calibrating how the feedforward gain is affected by the various factors and programming a gain-compensating table to these factors in the controller, it is feasible to realize a PDH feedforward system that is robust to environmental perturbations and external controls.

\section{Conclusion}

We present a simple yet effective method which greatly enhances the long-term stability of PDH feedforward. By constructing a feedforward circuitry whose gain dynamically adapts to the variations in cavity transmission power, we demonstrate a robust phase-noise suppression of over 40~dB under external perturbations. This adaptive-gain control technique can be readily extended to handle other types of perturbations not related to power fluctuations. The simplicity and the reliability demonstrated by the current technique may make PDH feedforward a routine solution for suppressing high-frequency laser phase noise. The method is expected to find immediate applications in enhancing coherent synthesis of ultracold molecules~\cite{2024_Cornish_STIRAP} and in improving fidelity of atomic quantum gates~\cite{2015_Schmidt_FrequencyComb,2018_Levine_highFidelity,2022_Senko_Limits,2023_Saffman_GateFidelity}.

\begin{acknowledgments}
This work is supported by the Innovation Program for Quantum Science and Technology (2021ZD0302104), and the National Natural Science Foundation of China (NSFC) (Grant No. 1223401, No. W2431002 and No. 92265205).
\end{acknowledgments}

\appendix
\section{Detailed experimental setup}\label{appendix_setup_details}
\begin{figure}[htbp]
	\centering
	\includegraphics[width=0.98\columnwidth]{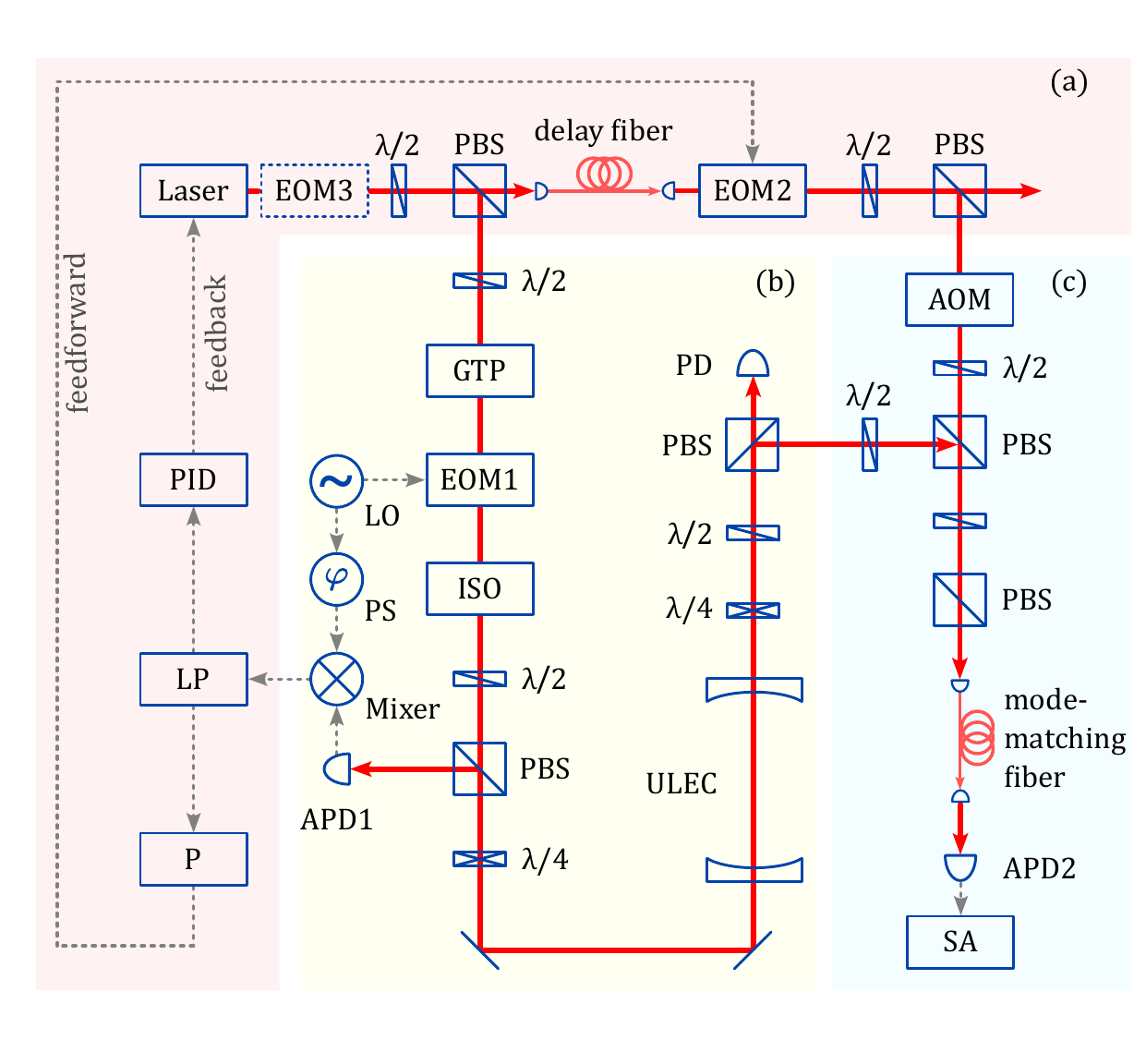}
	\caption{
	    Detailed schematic of our experimental setup. Part (a) shows the feedback and feedforward control systems; part (b) works to generate the PDH error signal; part (c) is the setup for measuring the self-heterodyne spectrum. EOM, electro-optic modulator for phase modulation. PBS, polarizing beam splitter. LP, low-pass filter. PID, loop filter whose output is the sum of proportion, integral, and derivative of the input signal. P, loop filter whose output is proportional to the input signal. GTP, Glan-Taylor polarizer. ISO, Faraday isolator. LO, local oscillator. PS, phase shifter. PD, photodiode. APD, avalanche photodiode. ULEC, cavity made of ultra-low expansion glass. AOM, acousto-optic modulator. SA, RF spectrum analyzer.
	}
	\label{fig_DetailedSetup}
\end{figure}

Figure~\ref{fig_DetailedSetup} reveals more details of our experimental setup. The light source is a 1012-nm fiber laser (Preci-Laser FL-SF-1012-S). PDH locking is implemented with a high-finesse cavity with a FWHM linewidth of 14.5~kHz using an incident power of 80~\si{\micro\watt} before the cavity. A home-made electro-optic modulator (EOM1) is used to produce a 79~MHz phase modulation with a modulation index $\beta \sim 1.08$. The light reflected from the cavity is detected by an avalanche photodiode (APD1, Menlo Systems APD210). The output signal of APD1 is demodulated by the same 79~MHz signal using a level-13 mixer (Minicircuits, ZX05-1MHW-S+). After demodulation, a low-pass filter (LP, Minicircuits SLP-50+) eliminates high-frequency ($\gtrsim 48$~MHz) noise from the error signal. One half of PDH error signal is sent to a voltage-controlled oscillator (VCO) which drives a double-pass AOM to lock the laser frequency to the cavity resonance; the other half is fed forward to a fiber EOM (EOM2, JENOPTIK PM1064) after passing through the loop filter P with a flat adaptive gain.

The heterodyne spectra in Fig.~\ref{Bandwidth}(a,b) in the main text are measured using the setup shown in Fig.~\ref{fig_DetailedSetup}(c). An AOM is used to shift the frequency of the feedforwarded light after it passes through EOM2. This shifted light is then combined with the beam transmitted through the cavity using a polarizing beam splitter (PBS). To facilitate beating between the two orthogonally polarized light after the PBS, we rotate their polarization with a half-wave plate and then use another PBS to pick out the field components with the common polarization. After that, both beams are coupled into a single-mode fiber to ensure perfect mode matching. The beat signal is detected by another APD (APD2, Thorlabs APD430), whose output is sent to a spectrum analyzer (SA, Rohde $\&$ Schwarz FSV3000). The powers of the cavity-filtered light and the feedforwarded light are about 3~\si{\micro\watt} and 25~\si{\micro\watt} before APD2, respectively. The beat note power of the two carriers is about 0~dBm.

To obtain the data shown in Fig.~\ref{Bandwidth}, we place another EOM (EOM3, KEYANG PHOTONICS KY-PM-10-10G) after the laser source (Fig.~\ref{fig_DetailedSetup}(a)) to inject a weak phase modulation ($<-40$~dBc), thereby imitating phase noise of variable frequency.

\section{Design details of loop-filter P}\label{appendix_circuit_details}

\begin{figure*}[ht]
	\centering
	\includegraphics[width=1.9\columnwidth]{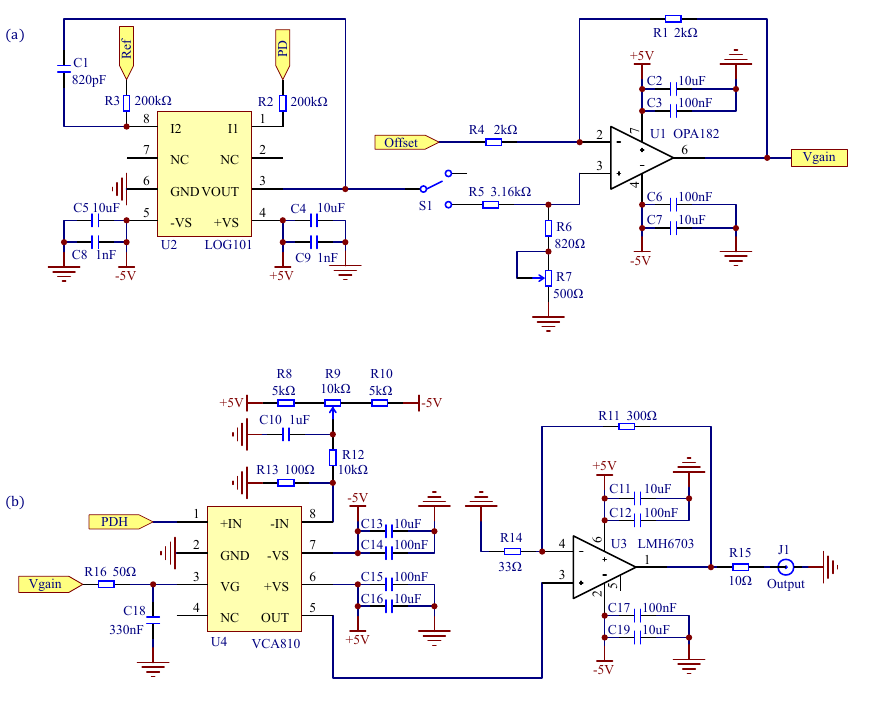}
	\caption{
		The essential parts of loop filter P that enables adaptive-gain control. (a) Circuit for producing gain-controlling voltage. Net ports labeled `PD', `Ref' and `Offset' correspond to $V_{\rm PD}(t)$, $V_{\rm ref}$, $V_{\rm offset}$ described in the main text. (b) The circuit that amplifies $V_{\rm PDH}(t)$ for feedforward, with adjustable gain controlled by $V_{\rm PDH}$. The net port `PDH' corresponds to the input signal $V_{\rm PDH}(t)$. The output port J1 is connected to EOM2 through a coaxial cable whose length is adjusted to match $t$ and $t^\prime$.
	}
	\label{fig_Circuitry}
\end{figure*}

The loop filter P contains two primary parts: one for generating the gain-controlling voltage (Fig.~\ref{fig_Circuitry}(a)) and another for amplifying $V_{\rm PDH}(t)$ with a variable gain (Fig.~\ref{fig_Circuitry}(b)). The first part aims to produce a gain which is inversely proportional to $V_{\rm PD}$, thereby compensating for slow drifts (less than a few kHz) in $P(t)$. For reliable performance, resistors with high precision and low temperature drift are used. The LOG101 logarithmic amplifier has output nonlinearity that depends on the input currents. Therefore, R3 and R4 are chosen to limit the input currents to an optimal region. The potentiometer R7 is used to fine-tune the value of $\kappa_1$. In addition to the variable gain amplifier VCA810, a wide-band operational amplifier LMH6703 is employed to provide the required amplification. To improve the feedforward bandwidth, the flatness of the gain and phase shift of all operational amplifiers in Fig.~\ref{fig_Circuitry}(b) are carefully considered. In addition, we find that the length of the coaxial cable between R15 and EOM2 has a notable influence on the flatness of the gain of loop filter P~\cite{cable_length_comment}. Consequently, the output resistor R15 is selected to mitigate this problem. To minimize nonlinear effects due to saturation, we recommend to select amplifiers with suitable saturation levels, adopt mixers with high LO level, and avoid using light power close to saturation level of APD1.

\section{Limitations of PDH Feedforward}\label{appendix_limitations}
\begin{figure}[ht]
	\centering
	\includegraphics[width=0.98\columnwidth]{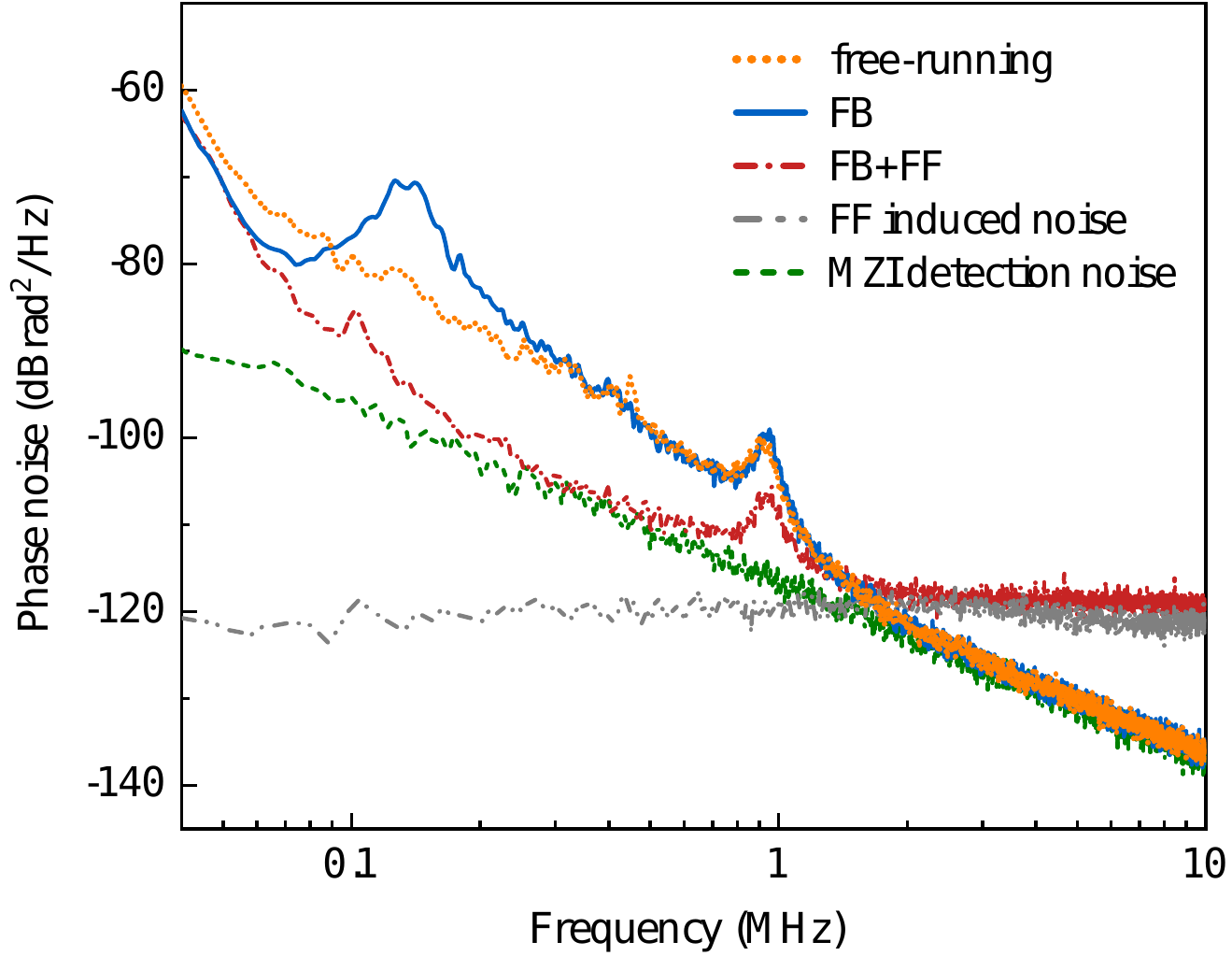}
	\caption{
		Single sideband power spectra of the laser's noise measured under different conditions using a delayed self-homodyne setup. See text for detailed discussions. 
	}
	\label{fig_PSD_PhaseNoise}
\end{figure}

It should be emphasized that the phase-noise attenuation presented in Fig.~\ref{Bandwidth}(c,d), which is calibrated by imprinting a strong phase modulation into the laser field, represents the optimal performance achievable by PDH feedforward for this particular design. Nonetheless, actual implementation effects on the laser’s phase noise may be constrained by various factors, which must be considered to approach optimal performance. 
 
To illustrate, Fig.\ref{fig_PSD_PhaseNoise} presents the power spectra of laser phase noise under three conditions: free-running (orange dotted line), locked with PDH feedback only (blue solid line), and locked with PDH feedback plus PDH feedforward implemented (red dashed-dotted line). The noise spectra are obtained from delayed self-homodyne measurements~\cite{2009_Hossein_MZI,2012_Hossein_MZI,2016_Tetsuya_TrackingInterferometer,2017_Lintz_Note,2021_Polzik_HighFrequency} through a Mach-Zehnder interferometer (MZI) with an optical path difference of 1.46~meters between its arms. A piezo actuator stabilizes the relative arm length, locking it with a bandwidth of several tens of kHz to balance the output powers of the MZI. Phase noise at frequencies beyond this locking range creates power imbalances between the two outputs, detected via a differential photodetector (Thorlabs PDB425C) to produce the noise spectra shown in Fig.~\ref{fig_PSD_PhaseNoise}.

The particular laser system used in this demonstration (Preci-Laser FL-SF-1012-S) possesses a relatively narrow free-running technical linewidth of about 10~kHz, but it exhibits a significant amplitude noise around 920~kHz, which PDH feedforward cannot suppress. Furthermore, due to its intrinsic design, applying feedback control to the laser frequency introduces amplitude modulation. Therefore, to prevent such feedback-induced amplitude modulation from affecting our feedforward results in Fig.~\ref{Bandwidth}, we restricted the feedback bandwidth to about 140~kHz in this work. This restriction explains the enhanced phase noise near 140~kHz and partially accounts for the weaker noise suppression below 100~kHz when PDH feedback is applied (blue solid line in Fig.\ref{fig_PSD_PhaseNoise}).

When PDH feedback and feedforward are applied simultaneously, the phase noise of the laser is suppressed for noise frequencies below 1.3~MHz but enhanced for higher frequencies, in comparison to that of the free-running laser. To gain a quantitative understanding of this behavior, we measure the sensitivity limit of the homodyne setup due to the detection noise (optical shot noise plus electrical noise) of the differential photodetector (Thorlabs PDB425C). This limit is represented by the green dashed line in Fig.~\ref{fig_PSD_PhaseNoise}. Notably, above 2~MHz, the phase noise levels of both the free-running and feedback-locked lasers are below this detection-noise limit.At frequencies near the MZI locking bandwidth of several tens of kHz, the MZI’s detection sensitivity is expected to fall below the detection-noise limit due to noise introduced by arm length stabilization. This effect accounts for why feedforward noise suppression appears less effective than the cavity transmission limit (see reference~\cite{2024_Chao_PDHFeedForward} for a detailed discussion) for noise frequencies below 600~kHz. Achieving an independent calibration of the MZI's sensitivity, however, would require a laser with ultra-low phase and amplitude noise.

To identify the source of enhanced phase noise above 1.3~MHz induced by feedforward, we measure the electrical output noise of loop filter P with the entire feedforward system active, but with PDH feedback locking turned off. This configuration captures contributions from the optical shot noise and electrical noise within the PDH detection system, as well as the electrical noise of loop filter P, while excluding laser phase noise effects. The resulting data, converted to single-sideband phase noise using the voltage-to-phase coefficient of EOM2, is represented by the gray dashed-dotted-dotted line in Fig.~\ref{fig_PSD_PhaseNoise}. This noise remains approximately flat at -120~$\rm dBrad^2/Hz$ up to 40~MHz, decreasing beyond that frequency. Despite this, the power of the noise pedestal generated by feedforward (integrated up to 100~MHz) remains below $0.01\%$ of the total laser power.

In summary, the above observations reveal the following insights: (i) The ultimate effectiveness of PDH feedforward on a laser is constrained by the detection sensitivity of the PDH setup; (ii) PDH detection offers generally higher sensitivity than delayed self-homodyne at lower frequencies; (iii) PDH feedforward is incapable of suppressing laser amplitude noise; (iv) PDH feedforward inevitably introduces additional phase noise, especially at high frequencies. Besides choosing quieter electrical components, a balance between speed and induced noise should be tailored to the actual noise spectrum of a laser. For instance, for lasers with negligible phase noise above $\sim 10$~MHz, one may consider using slower but quieter chips, or simply insert a low-pass filter between the loop filter P and EOM2 to isolate the high-frequency electrical noise.


\begin{thebibliography}{57}
\bibitem [{How()}]{HowToMatchDelay}%
	\BibitemOpen
	\href@noop {} {}\bibinfo {note} {Experimentally, we match $t$ and $t^\prime$
		by first choosing a delay fiber longer than the estimated effective length of
		the optical and electrical paths combined (i.e. from the beam splitter after
		the laser to APD1, and then from the APD1 to EOM2), and then compensating for
		the extra fiber's length by inserting coaxial cables between loop filter P
		and EOM2.}\BibitemShut {Stop}%
\bibitem [{cab()}]{cable_length_comment}%
	\BibitemOpen
	\href@noop {} {}\bibinfo {note} {To maximize the feedforward bandwidth and
		its performance, we use a large PDH modulation frequency close to 79 MHz and
		select high-speed electronic components with gain and group delay as flat as
		possible over a large frequency range. To achieve good gain flatness, the
		delay fiber should not be unnecessarily long, otherwise a long coaxial cable
		is needed to be inserted between the loop filter P and EOM2. This would
		result in a large cable capacitance and adversely affect the gain flatness of
		the amplifier LMH6703. In this work, we use a delay fiber of 10 meters long,
		and a coaxial cable of about 70~cm between the loop filter P and
		EOM2.}\BibitemShut {Stop}%
\bibitem [{off()}]{offsetlocking}%
	\BibitemOpen
	\href@noop {} {}\bibinfo {note} {In many experiments, the laser is
		offset-locked to the cavity by locking to one of the phase-modulation
		sidebands. We find that changing this frequency offset would also affect the
		strength of the PDH signal even if the cavity transmission power is fixed.
		This subtle effect is pressumaby caused by nonlinear interplays between the
		beat signal at the offset frequency and that at the PDH modulation frequency
		in APD1.}\BibitemShut {Stop}%
\end{thebibliography}
%

\end{document}